\documentclass[a4paper, conference]{IEEEtran}
\IEEEoverridecommandlockouts
\usepackage{cite}
\usepackage{stfloats, float}
\usepackage{amsmath,amssymb,amsfonts}
\usepackage[ruled]{algorithm2e}
\usepackage{graphicx, subfigure}
\usepackage{textcomp}
\usepackage{xcolor}
\usepackage{enumitem}
\usepackage{flushend}

\def\BibTeX{{\rm B\kern-.05em{\sc i\kern-.025em b}\kern-.08em
    T\kern-.1667em\lower.7ex\hbox{E}\kern-.125emX}}

\makeatletter
\renewcommand{\maketag@@@}[1]{\hbox{\m@th\normalsize\normalfont#1}}%
\makeatother

\begin{document}

\title{An NLoS-based Enhanced Sensing Method for MmWave Communication System
}

\author{\IEEEauthorblockN{
        Shiwen~He\IEEEauthorrefmark{1}\IEEEauthorrefmark{2}\IEEEauthorrefmark{3},
        Kangli~Cai\IEEEauthorrefmark{1},
        Shiyue Huang\IEEEauthorrefmark{1},
        Zhenyu An\IEEEauthorrefmark{3},
        Wei Huang\IEEEauthorrefmark{4},
        Ning Gao\IEEEauthorrefmark{5}
}
    
    \IEEEauthorblockA{\IEEEauthorrefmark{1}\emph{The School of Computer Science and Engineering, Central South University, Changsha 410083, China.}}
    \IEEEauthorblockA{\IEEEauthorrefmark{2}\emph{The National Mobile Communications Research Laboratory, Southeast University, Nanjing 210096, China.}}
    \IEEEauthorblockA{\IEEEauthorrefmark{3}\emph{The Purple Mountain Laboratories, Nanjing 211111, China.}}
    \IEEEauthorblockA{\IEEEauthorrefmark{4}\emph{The School of Computer Science and Information Engineering, Hefei University of Technology, Hefei
            230601, China.}}
    \IEEEauthorblockA{\IEEEauthorrefmark{5}\emph{Department of Standardization, OPPO Research Institute, Beijing, 100020, China.}}
    
    Email: \{shiwen.he.hn, caikangli, huangsy\}@csu.edu.cn, anzhenyu@pmlabs.com.cn, huangwei@hfut.edu.cn, gaoning1@oppo.com
}

\maketitle

\begin{abstract}
The millimeter-wave (mmWave)-based Wi-Fi sensing technology has recently attracted extensive attention since it provides a possibility to realize higher sensing accuracy. However, current works mainly concentrate on sensing scenarios where the line-of-sight (LoS) path exists, which significantly limits their applications. To address the problem, we propose an enhanced mmWave sensing algorithm in the 3D non-line-of-sight environment (mm3NLoS), aiming to sense the direction and distance of the target when the LoS path is weak or blocked. Specifically, we first adopt the directional beam to estimate the azimuth/elevation angle of arrival (AoA) and angle of departure (AoD) of the reflection path. Then, the distance of the related path is measured by the fine timing measurement protocol. Finally, we transform the AoA and AoD of the multiple non-line-of-sight (NLoS) paths into the direction vector and then obtain the information of targets based on the geometric relationship. The simulation results demonstrate that mm3NLoS can achieve a centimeter-level error with a 2m spacing. Compared to the prior work, it can significantly reduce the performance degradation under the NLoS condition.
\end{abstract}


\section{Introduction}
In recent years, Wi-Fi has been widely deployed in most public and private spaces due to its simplicity, reliability, and flexibility. The extremely dense Wi-Fi devices not only provide convenience for people, but also create a perfect opportunity to sense the environment. Therefore, by extracting appropriate signal features of Wi-Fi signals, e.g., phase differences\cite{PhaseBeat} or doppler shifts\cite{Widar}, we can effectively detect the presence of targets and further track them.

Target sensing based on Wi-Fi signals has been widely studied for lower frequencies, e.g., the fingerprint-based\cite{SShi} and geometry-based methods\cite{mDTrack}. These works achieved considerable performance due to the rich multipath signals in the environment and their weak attenuation characteristics. However, they critically depended on the channel state information (CSI), and the accuracy was limited by the antenna numbers and bandwidth. Moreover, these systems were designed for communication and did not consider the sensing function. To this end, the IEEE 802.11bf task group (TGbf) is working on making appropriate modifications to the Wi-Fi standard to utilize the existing 802.11-compatible waveforms for Wi-Fi sensing or integrated sensing and communication (ISAC)\cite{RDu}. Specifically, IEEE 802.11bf defines the support of 802.11ad and 802.11ay protocols, which significantly operate in the millimeter wave (mmWave) band. Therefore, a higher sensing performance can be expected in the future.
 
Although mmWave sensing is attractive, the short wavelength of mmWave leads to high path loss, and the propagation path is easily blocked by obstacles. To compensate for the attenuation, phased-array antennas and beamforming techniques for directional transmission are usually adopted. It means that one can estimate the angle of departure (AoD) and angle of arrival (AoA) from the directionally transmitted and received signals. Besides, the large bandwidth of mmWave provides a high distance resolution. Therefore, it is possible to realize accurate target sensing geometrically.

Prior work has demonstrated that mmWave could provide sub-decimeter accuracy in short-range sensing scenes, such as gesture tracking\cite{WiTrace}, mainly realized by leveraging two Wi-Fi links to detect the phase changes of CSI values due to the variation of propagation path length. However,  the transceivers significantly required a specific placement. The authors of \cite{polar} proposed a passive target sensing algorithm POLAR for IEEE 802.11ad devices, which used the AoD and time of flight (ToF) of the multi-path components estimated from channel impulse response (CIR)  corresponding to different beam patterns to sense the target. Still, it could only locate the object in 2D space. Furthermore, these systems were significantly designed based on the premise that the line-of-sight (LoS) path always existed. For non-line-of-sight (NLoS) conditions, the propagations of the signals were significantly affected, e.g., increased ToF and changed AoA. If the NLoS measurements were utilized directly as the LoS measurement, it would result in a large sensing error\cite{BHu}. To address this problem, the monostatic radar for sensing was proposed in \cite{mID}, which could directly estimate the range and relative radial speed using the received echo signal. Nevertheless, it is more attractive to realize mmWave sensing that can be applied for multi-device scenes, since the multi-angle detection for the target can remarkably improve sensing accuracy.

Based on the above consideration, this paper explores a sensing method between two devices in the 3D space. It is more challenging compared to the previous approach. On the one hand, the baseline information is significantly unknown due to the blocked LoS path, which makes solving bistatic triangles infeasible. On the other hand, diverse targets potentially lie on different planes. The complex geometry between targets, transmitter, and receiver further increases the difficulty. To address the above challenges, we propose an enhanced mmWave sensing algorithm in the 3D NLoS environment (mm3NLoS), which tries to sense the target by exploiting the geometric relationship of multiple NLoS paths. The main contributions of this work can be summarized as follows:
\begin{itemize}
    \item We design an enhanced sensing approach in a 3D NLoS environment so as to mitigate performance degradation when the LoS path is weak or blocked.
    \item We introduce the projection operation to simplify the problem and derive an analytical expression about the direction and distance between the target and receiver with the AoD, AoA, and ToF of two propagation paths.
    \item We compare the proposed method with the POLAR algorithm using simulated data. The result shows that our method performs better regardless of whether the LoS path exists.
\end{itemize}

\section{System Model}
\begin{figure}[pt]
    \centering
    \includegraphics[width=\linewidth]{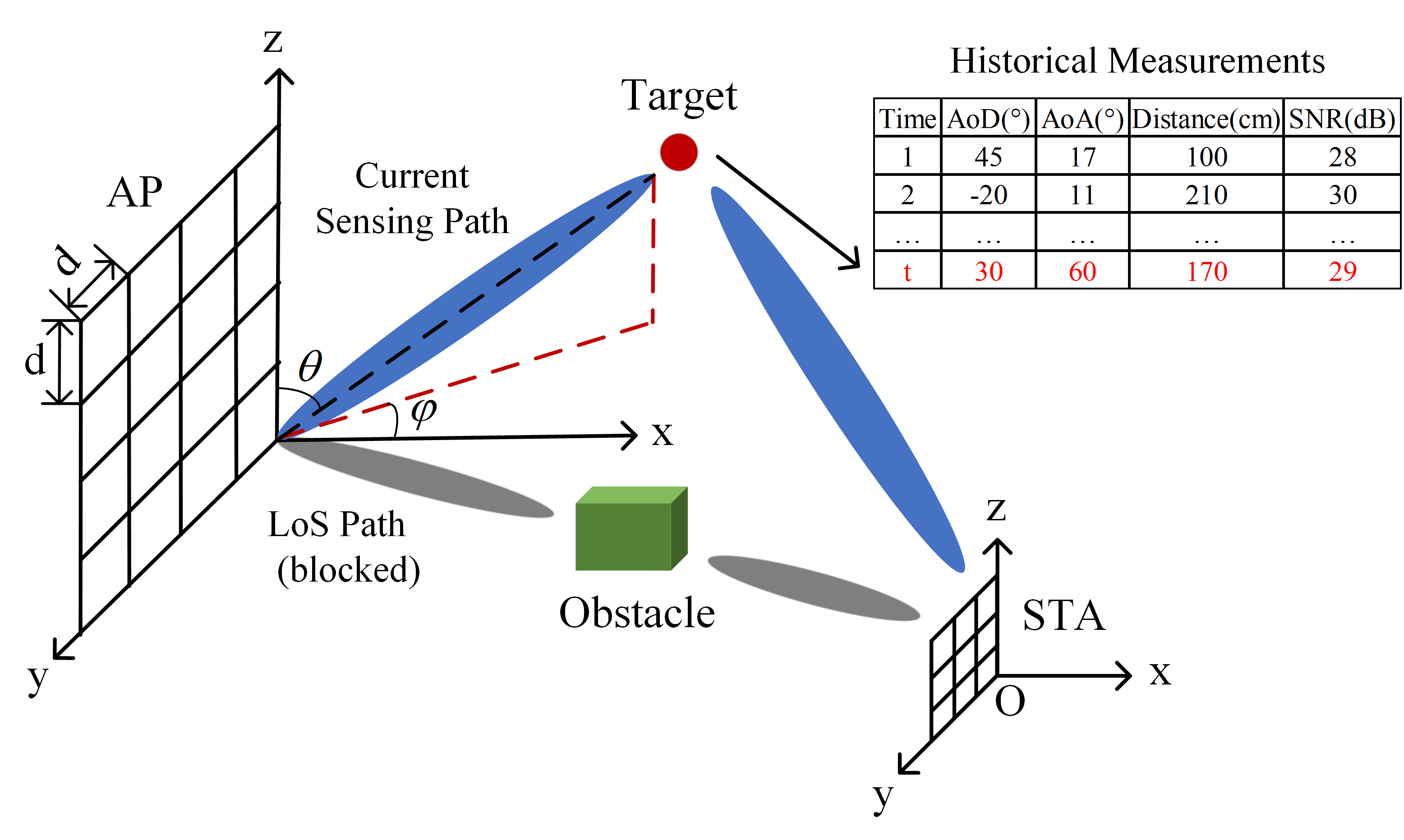}
    \caption{An illustration of target sensing in a 3D NLoS scene.}
    \label{fig:scene}
\end{figure}

In this paper, we exploit an mmWave MIMO system with an analog transceiver structure to sense a target. As shown in \figurename~\ref{fig:scene}, it consists of one access point (AP) and one station (STA), where AP and STA are the transmitter and receiver, respectively. We further assume that the LoS path is blocked by obstacles, so the NLoS path plays a dominant role in sensing. Generally, an NLoS path may be a multiple-bounce or single-bounce reflection. In this paper, we only consider the strongest propagation path and assume it to be single-bounce. In particular, utilizing multiple NLoS paths (the current and at least one historical path), the parameters of the target, including the distance and direction, can be estimated uniquely via geometric relations. To realize this, we record the AoD, AoA, ToF, and signal-to-noise ratio (SNR) of the sensing path in a historical measurement table and choose it based on the SNR.

Consider the uniform planar array (UPA) antennas and the single-path extended Saleh-Valenzuela geometric model for the mmWave system\cite{chen2018positioning}. Then, the channel matrix can be expressed as
\begin{equation}
    \mathbf{H} = \sqrt{N_tN_r}{g}{\mathbf{a}_r}({\varphi _{r}},{\theta _{r}}){\mathbf{a}_t^H}({\varphi _{t}},{\theta _{t}})\label{channelMatrix}
\end{equation}
where $g$ is the complex path gain with $g \sim \mathcal{CN}(0, 1)$. $N_t$ and $N_r$ are the antenna numbers of the AP and STA. $\varphi _{t}$ and $\theta _{t}$ are the azimuth and elevation of AoD. $\varphi _{r}$ and $\theta _{r}$ are the azimuth and elevation of AoA. For convenience, we assume that the UPA is placed in the YOZ plane, then the array response vectors are given by
\begin{equation}
    \begin{aligned}
        {{\bf{a}}_{t}}({\varphi_t},&{\theta_t}) = \frac{1}{{\sqrt {N_t} }}\left[ {1, \cdots ,{e^{ - jkd(p\sin {\varphi_t}\sin {\theta_t} + q\cos {\theta_t})}},}\right.
        \\[1mm]
        &\left. { \cdots ,{e^{ - jkd(({N_{t, h}} - 1)\sin {\varphi_t}\sin {\theta_t} + ({N_{t, v}} - 1)\cos {\theta_t})}}} \right]
    \end{aligned}
    \label{txArrayResponse}
\end{equation}
\begin{equation}
    \begin{aligned}
        {{\bf{a}}_r}({\varphi_r},&{\theta_r}) = \frac{1}{{\sqrt {N_r} }}\left[ {1, \cdots ,{e^{ - jkd(p\sin {\varphi_r}\sin {\theta_r} + q\cos {\theta_r})}},} \right.
        \\[1mm]
        &\left. { \cdots ,{e^{ - jkd(({N_{r, h}} - 1)\sin {\varphi_r}\sin {\theta_r} + ({N_{r, v}} - 1)\cos {\theta_r})}}} \right]
    \end{aligned}
    \label{rxArrayResponse}
\end{equation}
where $\lambda$ denotes the wavelength, and $k = \frac{2\pi}{\lambda}$. $N_{t, h}$ and $N_{t, v}$ respectively denote the numbers of transmit antennas in the horizontal and vertical directions, which satisfy $N_t = N_{t, h}N_{t, v}$. Similarly, $N_{r, h}$ and $N_{r, v}$ respectively denote the numbers of receive antennas in the horizontal and vertical directions, which satisfy $N_r = N_{r, h}N_{r, v}$. d is the inner-element spacing. $p$ and $q$ are the indices of elements, and $p = 0, 1, \dots, N_{a, h} - 1$, $q = 0, 1, \dots, N_{a, v} - 1$, $a \in \{t, r\}$.

Benefiting from the directional transmission and receive, the AoD and AoA can be obtained by a beam training procedure with a codebook. Specifically, the sensing signal is transmitted and received by different beams, in which the combination of the highest received SNR is used to estimate the angle information. For convenience, we assume that the $i$-th unit-norm codeword $\mathbf{f}_i$ and the $j$-th unit-norm codeword $\mathbf{w}_{j}$ of the Kronecker-Product codebook $\mathcal{C}$ are selected by AP and STA at time $t$, respectively. Then, the signal received by STA can be expressed as
\begin{equation}
    y_t = \sqrt{p_t} {\mathbf{w}_{j}^H}\mathbf{H}\mathbf{f}_{i}x_t + \mathbf{w}_{j}^H\mathbf{n}_t \label{receiveSignal}
\end{equation}
where $p_t$ and $x_t$ represent the transmit power and signal of AP at time $t$. $\mathbf{n}_t$ is the independent and identically distributed noise vector, and each element has $0$ mean and $\sigma^2$ variance. Then, the SNR can be calculated as
\begin{equation}
    \mathrm{SNR}_t = \frac{p_t{\left| {\mathbf{w}_j^H\mathbf{H}{\mathbf{f}_i}} \right|}^2}{\sigma ^2}
\end{equation}

After obtaining the AoA and AoD, the ToF of the propagation path could be further estimated by FTM with centimeter level errors\cite{pefkianakis2018accurate}. Then, utilizing the above-estimated parameters (i.e., AoD, AoA, and ToF) and the historical measurement information, we can locate the target (i.e., the direction and distance between the target and STA) with a geometrical method. Exceptionally, for the unavailable historical measurements (e.g., empty or same), we can replace them with other reflection paths, e.g., the first path, which is defined as the propagation path estimated to have the shortest ToF\cite{11ay}. To this end, a first path beamforming training similar to the 802.11ay protocol is required before the sensing.

\begin{figure*}[pt]
    \centering
    \subfigure[$\wp = 1$]{
        \includegraphics[width=0.31\linewidth]{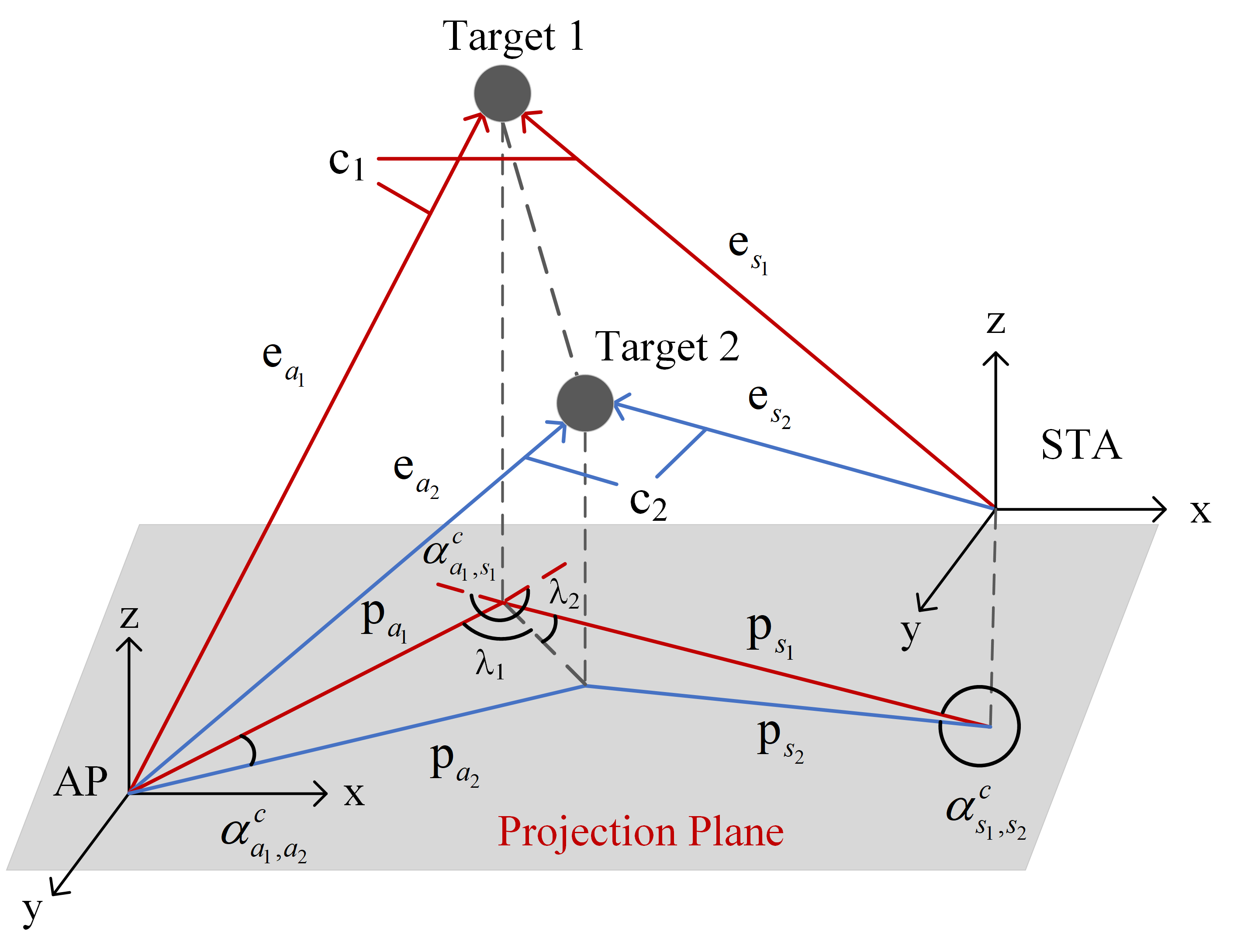}
        \label{sceneType1}
    }
    \subfigure[$\wp = 2$]{
        \includegraphics[width=0.31\linewidth]{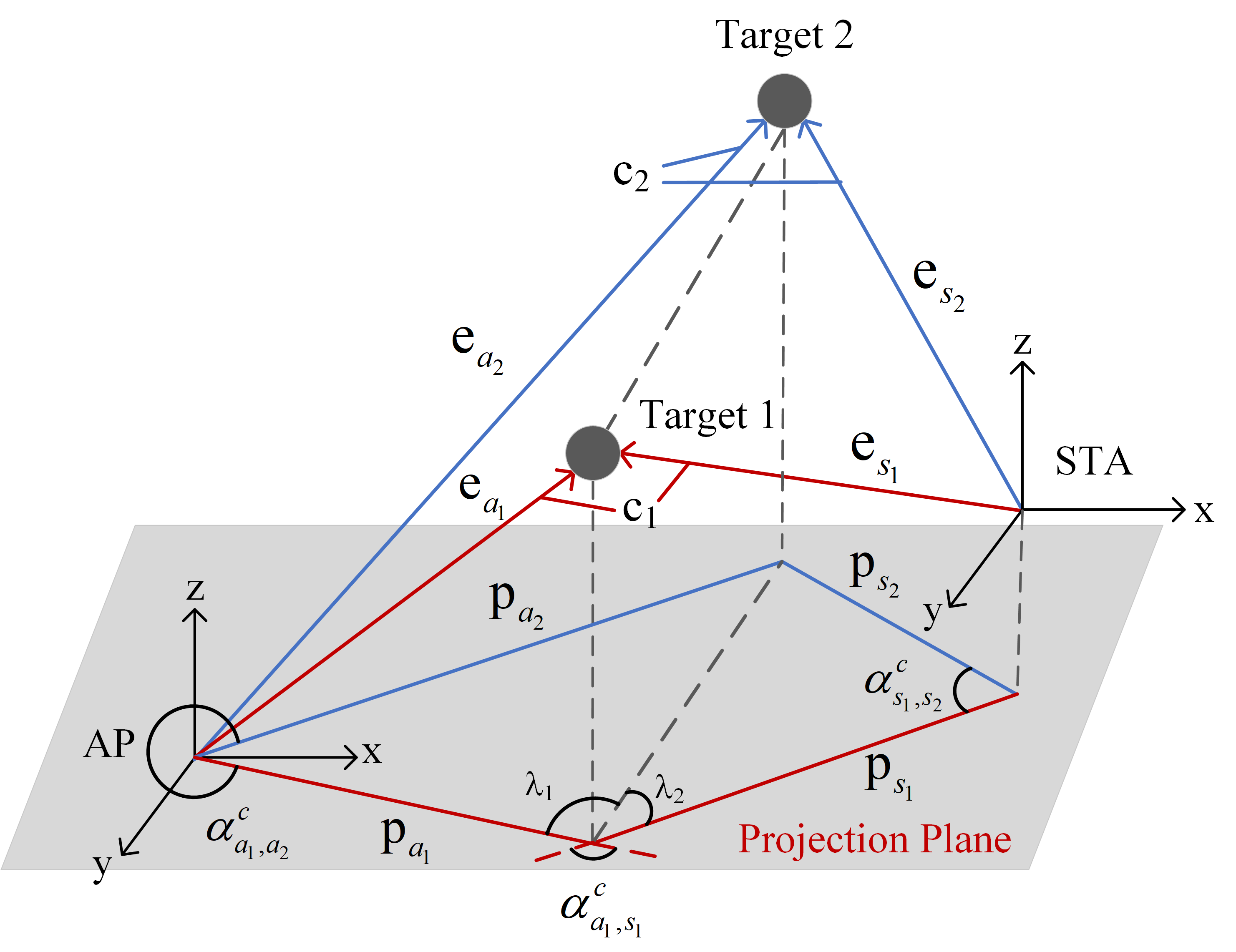}
        \label{sceneType2}
    }
    \subfigure[$\wp = 3$]{
        \includegraphics[width=0.31\linewidth]{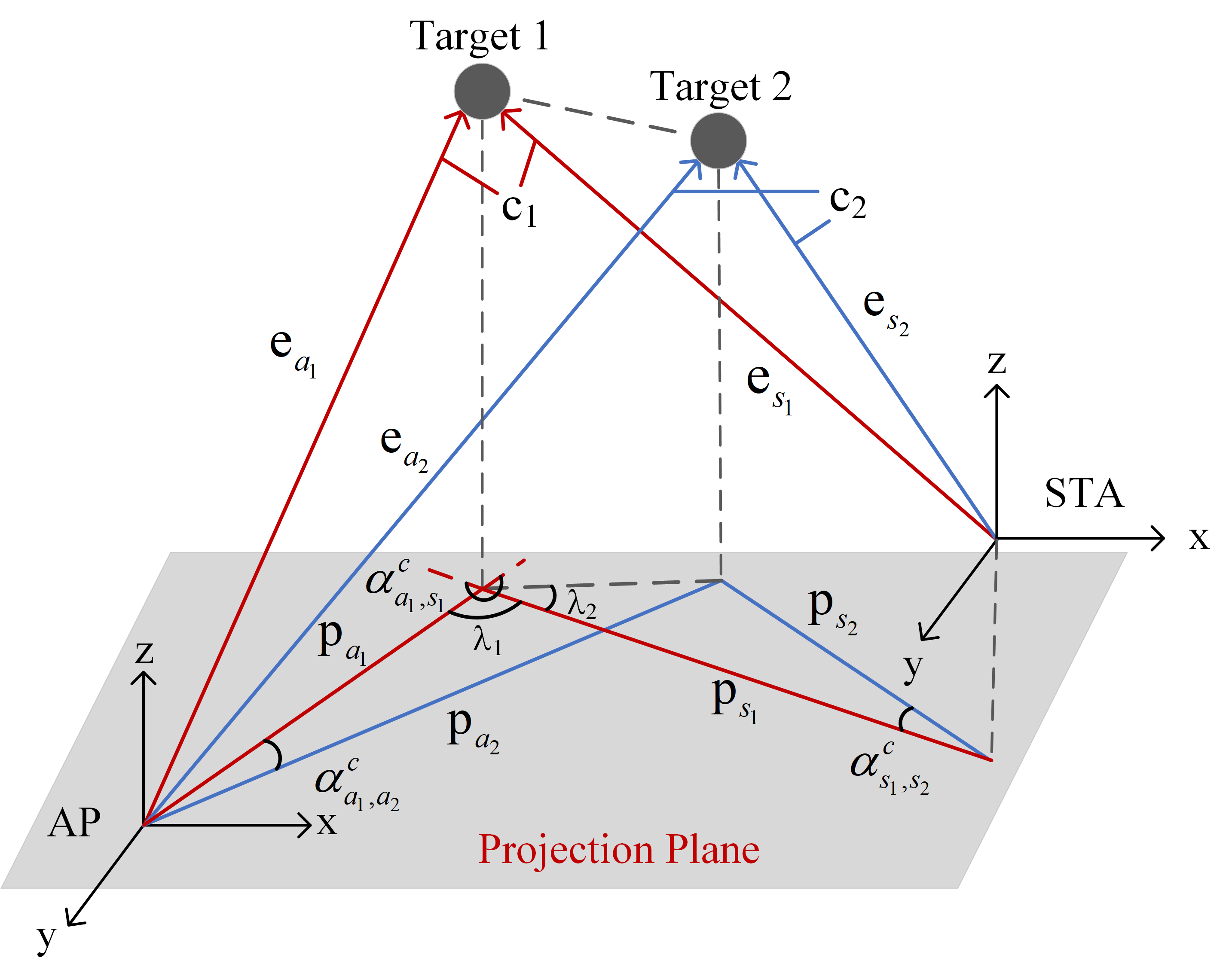}
        \label{sceneType3}
    }
    \subfigure[$\wp = 4$]{
        \includegraphics[width=0.31\linewidth]{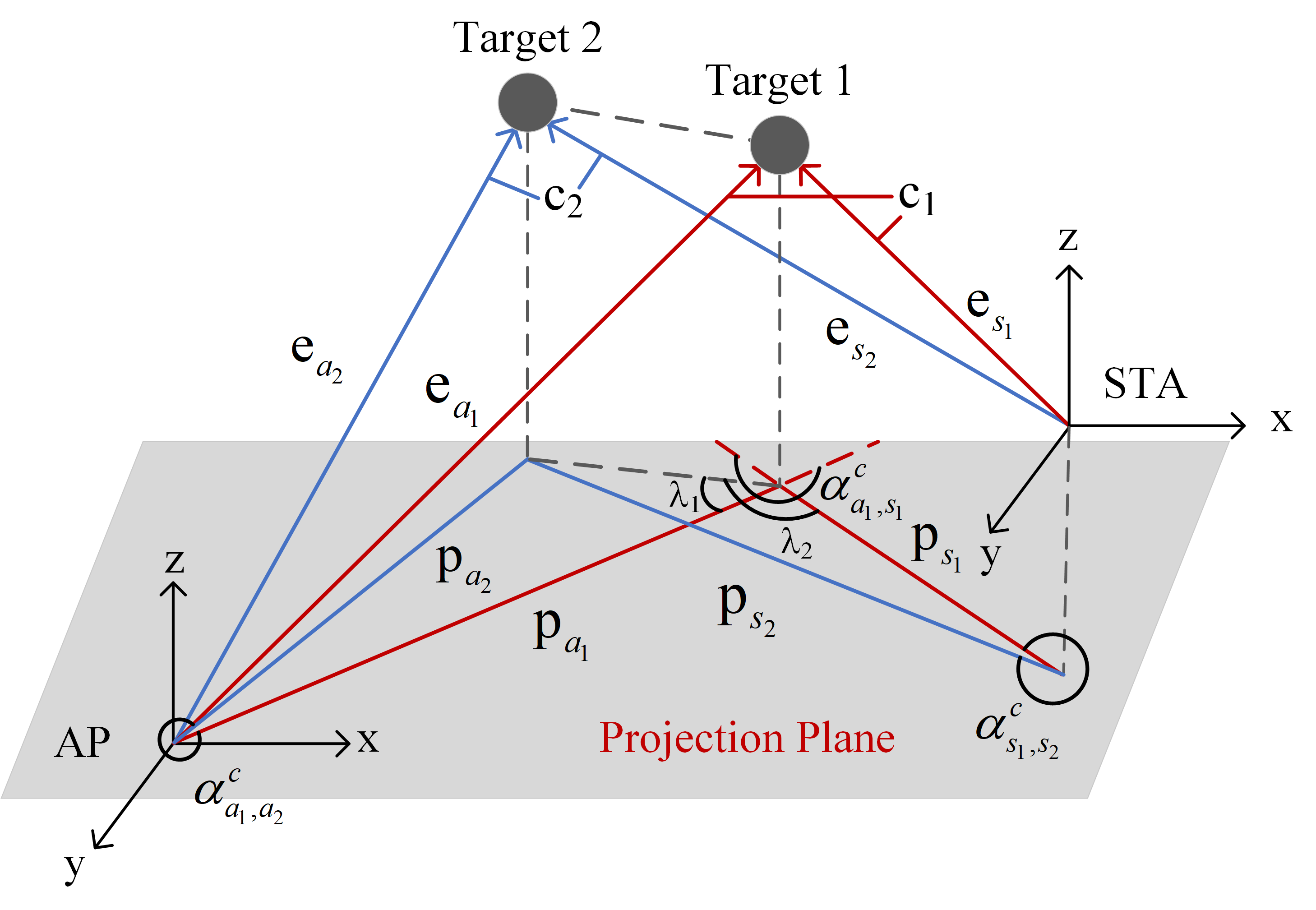}
        \label{sceneType4}
    }
    \subfigure[$\wp = 5$]{
        \includegraphics[width=0.31\linewidth]{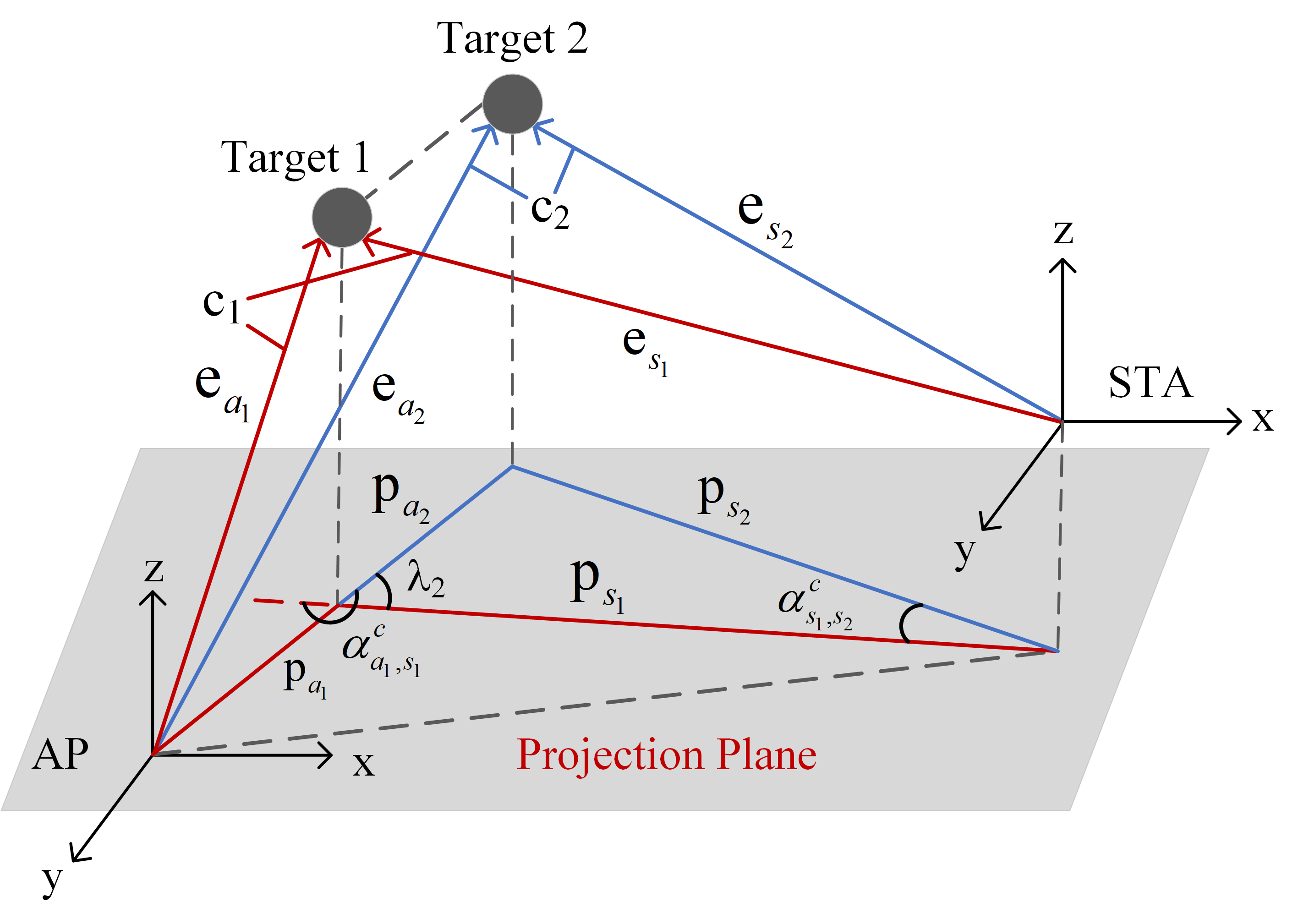}
        \label{sceneType5}
    }
    \caption{The possible projection cases.}
    \label{sceneTypes}
\end{figure*}

\section{Algorithm Design}
This section details the mm3NLoS, which realizes the sensing of the direction and the distance between the target and STA (i.e., $\mathbf{e}_{s_1}$ and $d_{s_1}$). Here, we only consider one historical NLoS path for simplicity. The sensing results are averaged if multiple historical NLoS paths are selected. Concretely,  we first convert the estimated AoD/AoA to a vector representation and project them into a 2D plane. Next, we classify the projection cases into five types according to the angle between the vectors. Finally, the distance $d_{s_1}$ between the target and STA is calculated for the given scene type.

\subsection{Scene Classification}
Since magnetometer has already been widely used in mobile devices, it is reasonable to assume that the AP and STA adopt the same reference orientation (e.g., x-axis) to measure the AoA and AoD. For ease of explanation, we denote the targets related to the current and selected historical NLoS path as Target 1 and Target 2, respectively. Then, we can transform the estimated AoD/AoA related to the target of different locations into a vector. For azimuth ${\varphi}_b$ and elevation ${\theta}_b$, the direction vector is calculated as
\begin{equation}
    \mathbf{e}_b = \left({{\rm{sin}}{\theta _{b}}{\rm{cos}}{\varphi _{b}}}, {{\rm{sin}}{\theta _{b}}{\rm{sin}}{\varphi _{b}}}, {{\rm{cos}}{\theta _{b}}}\right) \label{absoluteVector}
\end{equation}

The NLoS path related to Target $i \in \{1, 2\}$ can be described by $\mathbf{e}_{a_i}$, $\mathbf{e}_{s_i}$, and $c_i$, where $\mathbf{e}_{a_i}$ denotes the vector between AP and Target $i$, $\mathbf{e}_{s_i}$ represents the vector between STA and Target $i$, and $c_i$ is the NLoS path length calculated by ToF. It is extremely challenging to sense Target 1 due to the complex geometric relationship between vectors. To overcome the mentioned difficulties, we project the direction vector into a 2D plane to simplify the problem, as shown in \figurename~\ref{sceneTypes}. The projection vector can be given by
\begin{subequations}
    \begin{gather}
        {\mathbf{p}_{a_i}} = \mathbf{P}{\mathbf{e}_{a_i}}
        \\[1mm]
        {\mathbf{p}_{s_i}} = \mathbf{P}{\mathbf{e}_{s_i}}
    \end{gather}
\end{subequations}
where $\mathbf{P}$ is the projection matrix. For a given plane $\mathbf{B} = \left[\mathbf{b}_1\ \mathbf{b}_2\right]$, where $\mathbf{b}_1$ and $\mathbf{b}_2$ are the basis vectors, the projection matrix can be expressed by
\begin{equation}
    \mathbf{P} = \mathbf{B}{({\mathbf{B}^\top}\mathbf{B})^{ - 1}}{\mathbf{B}^\top} \label{projectionMatrix}
\end{equation}

With the projection vectors, we further classify them into different scene types, as shown in figure \figurename~\ref{sceneType1}$\sim$\ref{sceneType5}. Let ${\alpha}_{a_1,a_2}^c$, ${\alpha}_{s_1,s_2}^c$, and ${\alpha}_{a_1,s_1}^c$ respectively denote the clockwise angle between $\mathbf{p}_{a_1}$ and $\mathbf{p}_{a_2}$, $\mathbf{p}_{s_1}$ and $\mathbf{p}_{s_2}$, $\mathbf{p}_{a_1}$ and $\mathbf{p}_{s_1}$. Let $\mathcal{A}_1 = \left\{0, \pi, 2\pi\right\}$, $\mathcal{A}_2 = \left\{\alpha\left|\alpha \in \left(0, \pi\right)\right.\right\}$, and $\mathcal{A}_3 = \left\{\alpha\left|\alpha \in \left(\pi, 2\pi\right)\right.\right\}$, then, the scene type $\wp$ can be determined as follows.
\begin{description}
    \item[$\bullet$] $\wp = 0$, if $\alpha _{{a_1},{a_2}}^c, \alpha _{{s_1},{s_2}}^c \in \mathcal{A}_1$
    \item[$\bullet$] $\wp = 1$, if $\alpha _{{a_1},{a_2}}^c \in \mathcal{A}_2\; \mathrm{and}\; \alpha _{{s_1},{s_2}}^c \in \mathcal{A}_3$
    \item[$\bullet$] $\wp = 2$, if $\alpha _{{a_1},{a_2}}^c \in \mathcal{A}_3\; \mathrm{and}\; \alpha _{{s_1},{s_2}}^c \in \mathcal{A}_2$
    \item[$\bullet$] $\wp = 3$, if $\alpha _{{a_1},{a_2}}^c, \alpha _{{s_1},{s_2}}^c \in \mathcal{A}_2\; \mathrm{and}\; \alpha _{{a_1},{s_1}}^c \in \mathcal{A}_3$ or \\ $\alpha _{{a_1},{a_2}}^c, \alpha _{{s_1},{s_2}}^c \in \mathcal{A}_3\; \mathrm{and}\; \alpha _{{a_1},{s_1}}^c \in \mathcal{A}_2$
    \item[$\bullet$] $\wp = 4$, if $\alpha _{{a_1},{a_2}}^c, \alpha _{{s_1},{s_2}}^c, \alpha _{{a_1},{s_1}}^c \in \mathcal{A}_2$ or \\ $\alpha _{{a_1},{a_2}}^c, \alpha _{{s_1},{s_2}}^c, \alpha _{{a_1},{s_1}}^c \in \mathcal{A}_3$
    \item[$\bullet$] $\wp = 5$, if $\alpha _{{a_1},{a_2}}^c \in \mathcal{A}_1\; \mathrm{and}\; \alpha _{{s_1},{s_2}}^c \notin \mathcal{A}_1$ or \\ $\alpha _{{a_1},{a_2}}^c \notin \mathcal{A}_1\ \mathrm{and}\ \alpha _{{s_1},{s_2}}^c \in \mathcal{A}_1$
\end{description}
where $\wp = 0$ represents the unsolvable scene (i.e., AP, STA, Target 1, and Target 2 are collinear). It is worth mentioning that although there are so many scenes, most of them share a similar solution. Therefore, to simplify the procedure, we further classify the scene types into two categories, i.e., targets are separate ($\wp \in \{1, 2, 3, 4\}$) and targets are collinear ($\wp = 5$).

\begin{figure}[pt]
    \centering
    \includegraphics[width=0.8\linewidth]{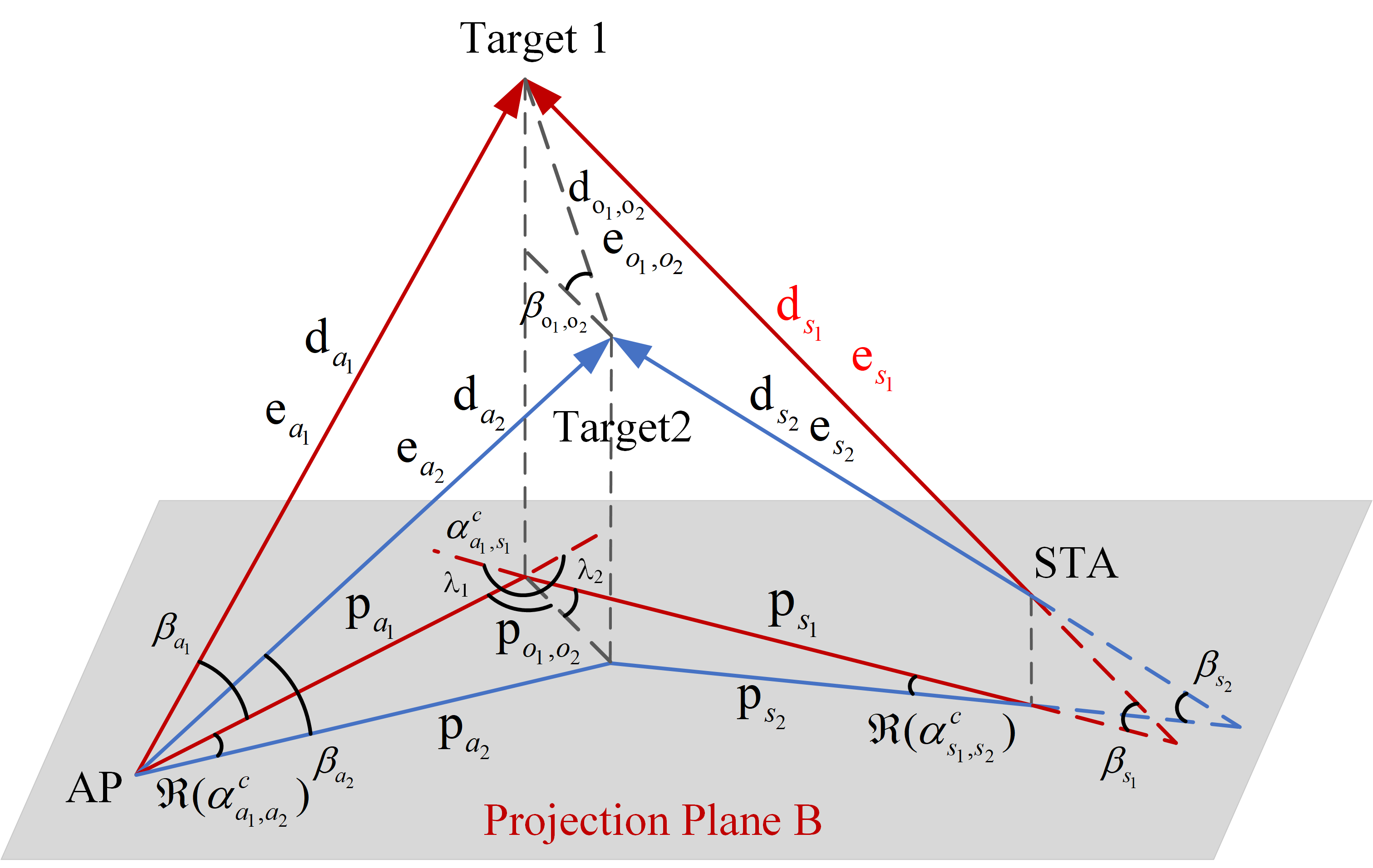}
    \caption{The geometric explanation when targets are separate ($\wp = 1$).}
    \label{fig:targetsSeparate}
\end{figure}

\begin{figure*}[bp]
    \hrulefill
    \vspace*{4pt}
    \setcounter{equation}{11}
    \begin{equation}
        \frac{{\sin (\Re (\alpha _{{a_1},{a_2}}^c) + {\lambda _1})\sin \Re (\alpha _{{s_1},{s_2}}^c)\cos {\beta _{{s_1}}} + \sin (\Re (\alpha _{{s_1},{s_2}}^c) + {\lambda _2})\sin \Re (\alpha _{{a_1},{a_2}}^c)\cos {\beta _{{a_1}}}}}{{\sin {\lambda _1}\sin \Re (\alpha _{{s_1},{s_2}}^c)\cos {\beta _{{s_2}}} + \sin {\lambda _2}\sin \Re (\alpha _{{a_1},{a_2}}^c)\cos {\beta _{{a_2}}}}} = \frac{{{c_1}\cos {\beta _{{a_1}}}\cos {\beta _{{s_1}}}}}{{{c_2}\cos {\beta _{{a_2}}}\cos {\beta _{{s_2}}}}} \label{simplifiedGeometryRelationship}
    \end{equation}
    \setcounter{equation}{14}
    \begin{equation}
        {\omega _1} = {\left[ {\begin{array}{*{20}{c}}
                    1\\[1mm]
                    {\delta _{{\lambda _1}}}\\[1mm]
                    -1\\[1mm]
                    {-\delta _{{\lambda _1}}}
            \end{array}} \right]^\top}\left[ {\begin{array}{*{20}{c}}
                {\cos \Re ({\alpha _{{a_1},{a_2}}})\sin \Re ({\alpha _{{s_1},{s_2}}})\cos {\beta _{{a_2}}}\cos {\beta _{{s_1}}}\cos {\beta _{{s_2}}}}\\[1mm]
                {\cos ({\lambda _2 -\delta_{\lambda_1}\lambda_1} + \Re ({\alpha _{{s_1},{s_2}}}))\sin \Re ({\alpha _{{a_1},{a_2}}})\cos {\beta _{{a_1}}}\cos {\beta _{{a_2}}}\cos {\beta _{{s_2}}}}\\[1mm]
                {{{{c_1}} \mathord{\left/
                            {\vphantom {{{c_1}} {{c_2}}}} \right.
                            \kern-\nulldelimiterspace} {{c_2}}}\sin \Re ({\alpha _{{s_1},{s_2}}})\cos {\beta _{{a_1}}}\cos {\beta _{{s_1}}}\cos {\beta _{{s_2}}}}\\[1mm]
                {{{{c_1}} \mathord{\left/
                            {\vphantom {{{c_1}} {{c_2}}}} \right.
                            \kern-\nulldelimiterspace} {{c_2}}}\cos ({\lambda_2-\delta_{\lambda_1}\lambda_1})\sin \Re ({\alpha _{{a_1},{a_2}}})\cos {\beta _{{a_1}}}\cos {\beta _{{a_2}}}\cos {\beta _{{s_1}}}}
        \end{array}} \right] \label{omega1}
    \end{equation}
    \begin{equation}
        {\omega _2} = {\left[ {\begin{array}{*{20}{c}}
                    1\\[1mm]
                    1\\[1mm]
                    -1
            \end{array}} \right]^\top}\left[ {\begin{array}{*{20}{c}}
                {\sin \Re ({\alpha _{{a_1},{a_2}}})\sin \Re ({\alpha _{{s_1},{s_2}}})\cos {\beta _{{a_2}}}\cos {\beta _{{s_1}}}\cos {\beta _{{s_2}}}}\\[1mm]
                {\sin ({\lambda _2-\delta_{\lambda_1}\lambda_1} + \Re ({\alpha _{{s_1},{s_2}}}))\sin \Re ({\alpha _{{a_1},{a_2}}})\cos {\beta _{{a_1}}}\cos {\beta _{{a_2}}}\cos {\beta _{{s_2}}}}\\[1mm]
                {{{{c_1}} \mathord{\left/
                            {\vphantom {{{c_1}} {{c_2}}}} \right.
                            \kern-\nulldelimiterspace} {{c_2}}}\sin {(\lambda _2-\delta_{\lambda_1}\lambda_1)}\sin \Re ({\alpha _{{a_1},{a_2}}})\cos {\beta _{{a_1}}}\cos {\beta _{{a_2}}}\cos {\beta _{{s_1}}}}
        \end{array}} \right] \label{omega2}
    \end{equation}
\end{figure*}

\subsection{Targets Are Separate}
In this subsection, we focus on obtaining the target's distance relative to the STA based on the vectors' angles and the propagation path length. As shown in \figurename~\ref{fig:targetsSeparate}, $\mathbf{e}_{o_1, o_2}$ and $\mathbf{p}_{o_1, o_2}$ denote the vector formed by Target 1 and Target 2 and their projection vector on plane $\mathbf{B}$, respectively. ${\beta}_{a_1}$, ${\beta}_{a_2}$, ${\beta}_{s_1}$, ${\beta}_{s_2}$, and ${\beta}_{o_1, o_2}$ denote the angle between $\mathbf{e}_{a_1}$ and $\mathbf{p}_{a_1}$, $\mathbf{e}_{a_2}$ and $\mathbf{p}_{a_2}$, $\mathbf{e}_{s_1}$ and $\mathbf{p}_{s_1}$, $\mathbf{e}_{s_2}$ and $\mathbf{p}_{s_2}$, $\mathbf{e}_{o_1, o_2}$ and $\mathbf{p}_{o_1, o_2}$. $\lambda_1$ and $\lambda_2$ denote the complementary angle between $\mathbf{p}_{o_1, o_2}$ and $\mathbf{p}_{a_1}$, $\mathbf{p}_{o_1, o_2}$ and $\mathbf{p}_{s_1}$. ${d_{a_i}}$, ${d_{s_i}}$, and ${d_{{o_1},{o_2}}}$ are the distance between AP and Target $i$, STA and Target $i$, Target 1 and Target 2. According to the geometric relations, we have the following equations

\setcounter{equation}{8}
\begin{subequations}\label{geometryRelationship}
    \small
    \begin{gather}
        \frac{{\sin \Re (\alpha _{{a_1},{a_2}}^c)}}{{{d_{{o_1},{o_2}}}}cos{\beta}_{o_1,o_2}} = \frac{{\sin {\lambda _1}}}{{{d_{a_2}}}cos{\beta}_{a_2}} = \frac{{\sin (\Re (\alpha _{{a_1},{a_2}}^c) + {\lambda _1})}}{{{d_{a_1}}}cos{\beta}_{a_1}} \label{sineFormulaInAP}
        \\[1mm]
        \frac{{\sin \Re (\alpha _{{s_1},{s_2}}^c)}}{{{d_{{o_1},{o_2}}}}cos{\beta}_{o_1,o_2}} = \frac{{\sin {\lambda _2}}}{{{d_{s_2}}}cos{\beta}_{s_2}} = \frac{{\sin (\Re (\alpha _{{s_1},{s_2}}^c) + {\lambda _2})}}{{{d_{s_1}}}cos{\beta}_{s_1}} \label{sineFormulaInSTA}
        \\[1mm]
        {d_{a_1}} + {d_{s_1}} = {c_1} \label{distanceAboutTarget1}
        \\[1mm]
        {d_{a_2}} + {d_{s_2}} = {c_2} \label{distanceAboutTarget2}
    \end{gather}
\end{subequations}
where $\Re(\alpha) = min(\alpha, 2\pi - \alpha)$. $d_{s_1}$ is the value to be estimated, $\lambda_1$, $\lambda_2$, $d_{o_1, o_2}$, $d_{a_1}$, $d_{a_2}$, $d_{s_2}$ and $\beta_{{o_1},{o_2}}$ are the unknown variables. Combining \eqref{sineFormulaInAP}, \eqref{sineFormulaInSTA}, and \eqref{distanceAboutTarget1}, we could rewrite $d_{s_1}$ as follows
\begin{equation}
    {d_{{s_1}}} = \frac{{{g_2}{c_1}}}{{{g_1} + {g_2}}} \label{distanceWhenSeparate}
\end{equation}
where $g_1$ and $g_2$ are respectively defined as
\begin{subequations}\label{gValueWhenSeparate}
    \begin{gather}
        {g_1} = \sin (\Re (\alpha _{{a_1},{a_2}}^c) + {\lambda _1})\sin \Re (\alpha _{{s_1},{s_2}}^c)\cos {\beta _{{s_1}}}
        \\[1mm]
        {g_2} = \sin (\Re (\alpha _{{s_1},{s_2}}^c) + {\lambda _2})\sin \Re (\alpha _{{a_1},{a_2}}^c)\cos {\beta _{{a_1}}}
    \end{gather}
\end{subequations}

From \eqref{distanceWhenSeparate} and \eqref{gValueWhenSeparate}, we can see that the value of $d_{s_1}$ is related to $\lambda_1$ and $\lambda_2$, which could be solved with the following steps. First, by substituting \eqref{sineFormulaInAP}, \eqref{sineFormulaInSTA} to \eqref{distanceAboutTarget1}, \eqref{distanceAboutTarget2} and eliminating $d_{o_1,o_2}cos{\beta}_{o_1,o_2}$ simultaneously, we have \eqref{simplifiedGeometryRelationship}. Besides, from \figurename~\ref{sceneType1}$\sim$\ref{sceneType4}, we have
\setcounter{equation}{12}
\begin{equation}
    {\lambda _2} = \left\{ {\begin{array}{*{20}{l}}
            {2\pi  - \alpha _{{a_1},{s_1}}^c - {\lambda _1},}\\[1mm]
            {\alpha _{{a_1},{s_1}}^c - {\lambda _1},}\\[1mm]
            { - \Re (\alpha _{{a_1},{s_1}}^c) + {\lambda _1},}\\[1mm]
            {\Re (\alpha _{{a_1},{s_1}}^c) + {\lambda _1},}
    \end{array}} \right.\begin{array}{*{20}{c}}
        {\wp = 1}\\[1mm]
        {\wp = 2}\\[1mm]
        {\wp = 3}\\[1mm]
        {\wp = 4}
    \end{array}
    \label{angleRelation}
\end{equation}

Afterwards, combining \eqref{simplifiedGeometryRelationship} and \eqref{angleRelation}, $\lambda_1$ can be given by
\begin{equation}
    {\lambda _1} = {\tan ^{ - 1}}\frac{{ - {\omega _2}}}{{{\omega _1}}}
\end{equation}
where $\omega_1$ and $\omega_2$ are calculated using \eqref{omega1} and \eqref{omega2}, respectively. $\delta _{{\lambda _1}}$ is assigned according to the sign of $\lambda_1$ in the \eqref{angleRelation}, i.e., $\delta _{{\lambda _1}} = -1$ if ${\wp \in \{1, 2\}}$, and $\delta _{{\lambda _1}} = 1$ if ${\wp \in \{3, 4\}}$. With the calculated $\lambda_1$ and $\lambda_2$, the distance between the STA and the targets can be estimated using \eqref{distanceWhenSeparate}.

\begin{figure}[pt]
    \centering
    \includegraphics[width=0.8\linewidth]{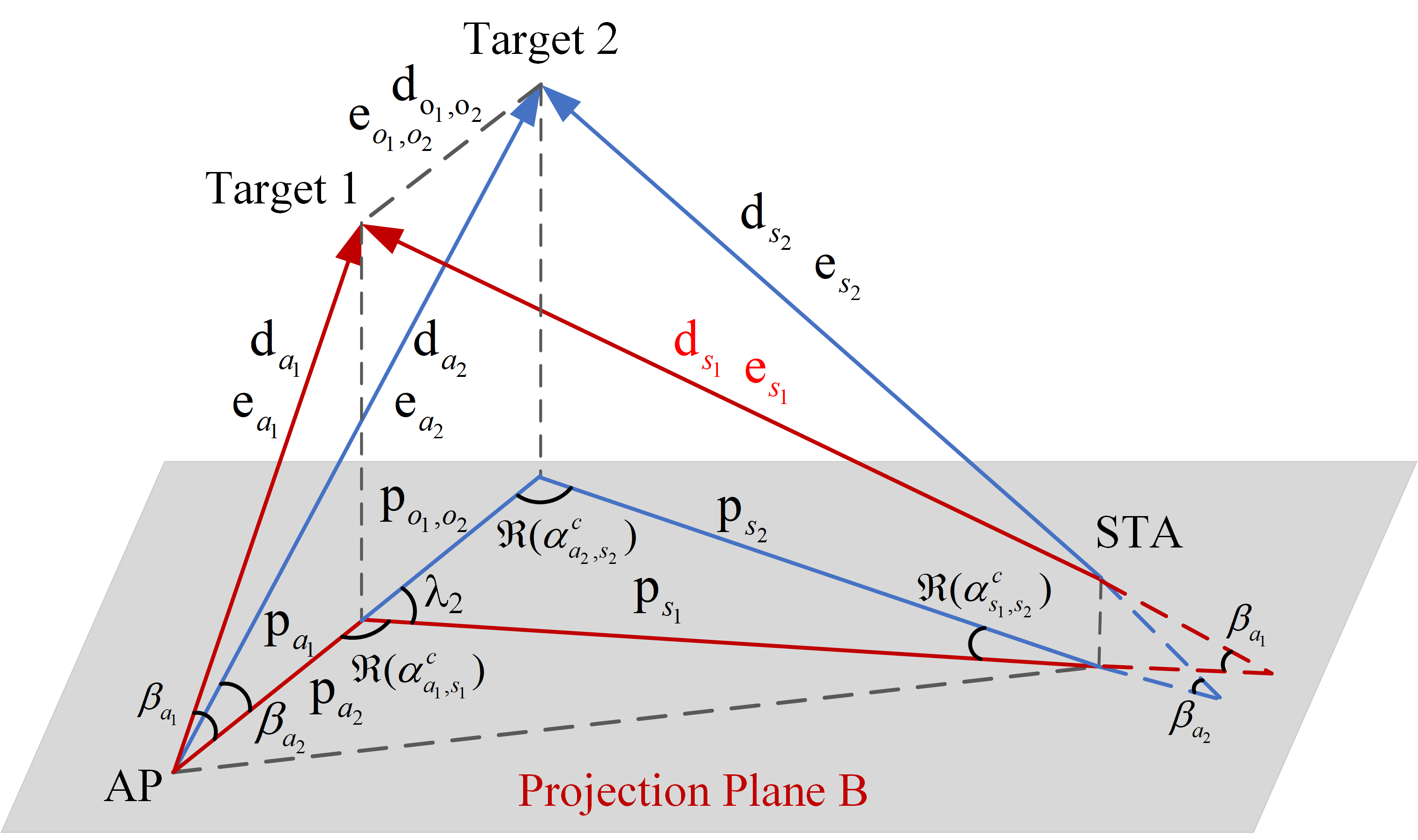}
    \caption{The geometric explanation when targets are collinear with AP.}
    \label{fig:targetsCollinear}
\end{figure}
 
\subsection{Targets Are Collinear}
When the collinear case occurs, $\lambda_1$ and $\lambda_2$ cannot simultaneously exist. Therefore, we need to consider the two cases individually (i.e., projections of targets are collinear with AP or STA). Here, we assume that they are collinear with AP for simplicity. As demonstrated in \figurename~\ref{fig:targetsCollinear}, we have

\setcounter{equation}{16}
\begin{subequations}
    \small
    \begin{gather}
        \frac{{\sin {\lambda _2}}}{{{d_{{s_2}}}\cos {\beta _{{s_2}}}}} = \frac{{\sin \Re (\alpha _{{s_1},{s_2}}^c)}}{{{d_{{o_1},{o_2}}}\cos {\beta _{{o_1},{o_2}}}}} = \frac{{\sin ({\lambda _2} + \Re (\alpha _{{s_1},{s_2}}^c))}}{{{d_{{s_1}}}\cos {\beta _{{s_1}}}}} \label{sineFormulaWhenCollinear}
        \\[1mm]
        {d_{{o_1},{o_2}}}\cos {\beta _{{o_1},{o_2}}} = {\delta _{{a_1}}}{d_{{a_1}}}\cos {\beta _{{a_1}}} + {\delta _{{a_2}}}{d_{{a_2}}}\cos {\beta _{{a_2}}} \label{distanceBetweenTargets}
    \end{gather}
\end{subequations}
where 
\begin{subequations}
    \small
    \begin{gather}
        {\delta _{{a_1}}} = \left\{ {\begin{array}{*{20}{l}}
                {1,}&{\Re (\alpha _{{a_1},{a_2}}^c) = 0,\Re (\alpha _{{a_1},{s_1}}^c) < \Re (\alpha _{{a_2},{s_2}}^c)}
                \\[1mm]
                { - 1,}&{\Re (\alpha _{{a_1},{a_2}}^c) = 0,\Re (\alpha _{{a_1},{s_1}}^c) > \Re (\alpha _{{a_2},{s_2}}^c)}
                \\[1mm]
                {1,}&{\Re (\alpha _{{a_1},{a_2}}^c) = \pi }
        \end{array}} \right. \\[1mm]
        {\delta _{{a_2}}} = \left\{ {\begin{array}{*{20}{l}}
                {-1,}&{\Re (\alpha _{{a_1},{a_2}}^c) = 0,\Re (\alpha _{{a_1},{s_1}}^c) < \Re (\alpha _{{a_2},{s_2}}^c)}
                \\[1mm]
                {1,}&{\Re (\alpha _{{a_1},{a_2}}^c) = 0,\Re (\alpha _{{a_1},{s_1}}^c) > \Re (\alpha _{{a_2},{s_2}}^c)}
                \\[1mm]
                {1,}&{\Re (\alpha _{{a_1},{a_2}}^c) = \pi }
        \end{array}} \right.
    \end{gather}
\end{subequations}

Then, based on \eqref{distanceAboutTarget1}, \eqref{distanceAboutTarget2}, \eqref{sineFormulaWhenCollinear}, and \eqref{distanceBetweenTargets}, we can obtain the distance between the STA and Target 1 as follows
\begin{equation}
    {d_{{s_1}}} = \frac{{\sin ({\lambda _2} + \Re (\alpha _{{s_1},{s_2}}^c))\cos {\beta _{{s_2}}}}}{{\sin {\lambda _2}\cos {\beta _{{s_1}}}}} \times ({c_2} - \frac{{{g_1}}}{{{g_2}}}) \label{distanceWhenCollinear}
\end{equation}
where
\begin{subequations}
    \small
    \begin{gather}
        {g_1} = {\left[ {\begin{array}{*{20}{c}}
                    {{c_2}}\\[1mm]
                    {{\delta _{{a_1}}}{c_2}}\\[1mm]
                    {{\delta _{{a_1}}}{c_1}}
            \end{array}} \right]^\top}\left[ {\begin{array}{*{20}{c}}
                {\sin \Re (\alpha _{{s_1},{s_2}}^c)\cos {\beta _{{s_1}}}\cos {\beta _{{s_2}}}}\\[1mm]
                {\sin ({\lambda _2} + \Re (\alpha _{{s_1},{s_2}}^c))\cos {\beta _{{a_1}}}\cos {\beta _{{s_2}}}}\\[1mm]
                {-\sin {\lambda _2}\cos {\beta _{{a_1}}}\cos {\beta _{{s_1}}}}
        \end{array}} \right]
        \\[1mm]
        {g_2} = {\left[ {\begin{array}{*{20}{c}}
                    1\\[1mm]
                    {{\delta _{{a_1}}}}\\[1mm]
                    {{\delta _{{a_2}}}}
            \end{array}} \right]^\top}\left[ {\begin{array}{*{20}{c}}
                {\sin \Re (\alpha _{{s_1},{s_2}}^c)\cos {\beta _{{s_1}}}\cos {\beta _{{s_2}}}}\\[1mm]
                {\sin ({\lambda _2} + \Re (\alpha _{{s_1},{s_2}}^c))\cos {\beta _{{a_1}}}\cos {\beta _{{s_2}}}}\\[1mm]
                {\sin {\lambda _2}\cos {\beta _{{a_2}}}\cos {\beta _{{s_1}}}}
        \end{array}} \right]
    \end{gather}
\end{subequations}
It is obvious that the value of $d_{s_1}$ only depends on $\lambda_2$, which is calculated by
\begin{equation}
    \small
    {\lambda _2} = \left\{ {\begin{array}{*{20}{l}}
            {\Re (\alpha _{{a_1},{s_1}}^c),}&{\Re (\alpha _{{a_1},{a_2}}^c) = 0,\Re (\alpha _{{a_1},{s_1}}^c) < \Re (\alpha _{{a_2},{s_2}}^c)}\\[1mm]
            {\pi  - \Re (\alpha _{{a_1},{s_1}}^c),}&{\Re (\alpha _{{a_1},{a_2}}^c) = 0,\Re (\alpha _{{a_1},{s_1}}^c) > \Re (\alpha _{{a_2},{s_2}}^c)}\\[1mm]
            {\Re (\alpha _{{a_1},{s_1}}^c),}&{\Re (\alpha _{{a_1},{a_2}}^c) = \pi }
    \end{array}} \right.\label{lambda2Formula}
\end{equation}
With the estimated $\lambda_2$, the distance between Target 1 and STA is determined uniquely.

It is worth noting that for a special case (i.e., Target 2 is on the LoS path), the problem will change into an LoS situation, and Target 2 is virtual at this time. Therefore, mm3NLoS is still effective for one LoS and one NLoS path. It can also be proved by \eqref{distanceWhenSeparate} and \eqref{distanceWhenCollinear}, in which the calculation result is only related to the angles between vectors and the distance of each propagation path, and the LoS path can be regarded as the two vectors on the same line. Also, in practice, one sees that the angle estimation is sensitive to array size, which may make mm3NLoS underperform for the small antenna array. To address this issue, we adopt the auxiliary beam pair to refine the AoD and AoA\cite{DalinZhu}, which is realized by deflecting the best beam a small phase in the horizontal and vertical domains, respectively.

\section{SIMULATION RESULTS}
In this section, we evaluate the performance of the proposed mm3NLoS algorithm. We consider an mmWave MIMO system in the 60 GHz band. AP and STA employ the UPA with the vertical and horizontal spacing of $\lambda / 2$ and assume that they can cover the $120^{\circ}$ and $90^{\circ}$ angular ranges in the azimuth and elevation directions, respectively. Besides, we set the projection plane as the YOZ plane. For convenience, the AP and STA are placed at $(0, 0, 0)$ and $(2, 0, 0)$, respectively.

\begin{figure}[pt]
    \centering
    \includegraphics[width=0.85\linewidth]{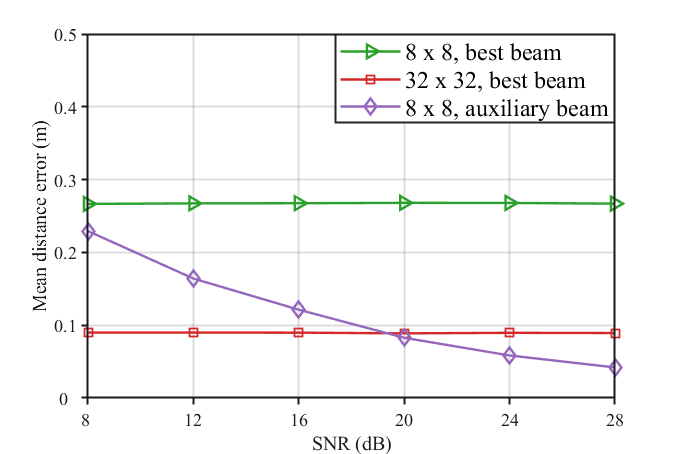}
    \caption{The impact of antenna numbers.}
    \label{fig:angle}
\end{figure}

We first measure the mean distance error of different antenna numbers for a given FTM measure variance with $\rho = 0.01$. The mean distance error is defined as the average distance between the actual location and the location calculated by estimated direction and distance in 2000 experiments. To mitigate performance degradation caused by the small antenna array size, we also adopt auxiliary beams to refine the AoD/AoA\cite{DalinZhu}. As demonstrated in \figurename~\ref{fig:angle}, the mean distance error drops with the increase of the antenna numbers. It is consistent with our intuition because more antennas generate a narrower beam pattern, thus better focusing on the reflection path. Furthermore, we observe that the mean distance error of auxiliary beams is significantly related to SNR, which is even below $32 \times 32$ UPA when SNR is greater than 20 dB. It can be explained that, for a higher SNR, the measurement of the auxiliary beam is less affected by noise, thus achieving a more accurate AoD and AoA estimation.

\begin{figure*}
    \begin{minipage}{0.5\linewidth}
        \centering
        \includegraphics[width=0.85\linewidth]{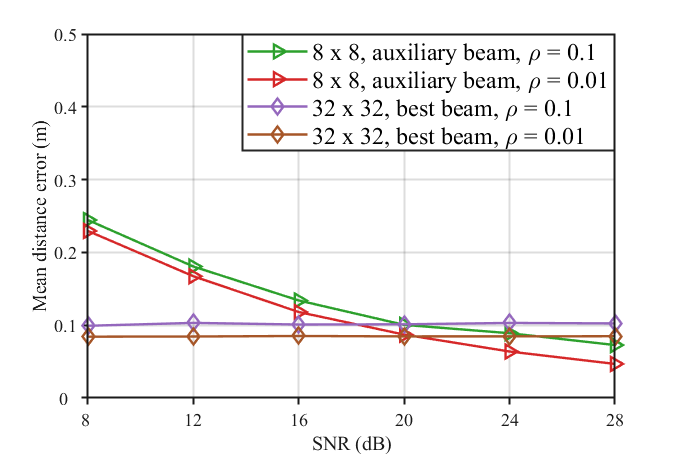}
        \caption{The impact of FTM variances.}
        \label{fig:rho}
    \end{minipage}
    \begin{minipage}{0.5\linewidth}
        \centering
        \includegraphics[width=0.85\linewidth]{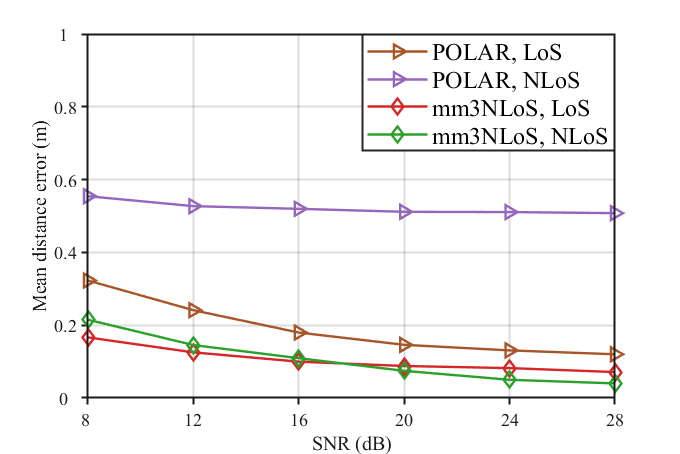}
        \caption{mm3NLoS vs. POLAR.}
        \label{fig:compare}
    \end{minipage}
\end{figure*}

Secondly, we further validate the impact of FTM measure variance $\rho^2$ when $32 \times 32$ UPA with the best beam and $8 \times 8$ UPA with auxiliary beams are adopted, respectively. There are two critical results worth noting in \figurename~\ref{fig:rho}. On the one hand, the error decreases with a smaller FTM measure variance for best beam measurements and is significantly independent of SNR. The reason is that the error is jointly caused by inaccurate distance and AoD/AoA estimation, in which the latter remains constant for the same best beam under all SNRs. On the other hand, we see an increased effect of FTM measure variances on the error for auxiliary beam measurements since more accurate AoD/AoA could be obtained in the case of higher SNR.

Finally, we compare mm3NLoS with POLAR proposed in \cite{polar}. For simplicity, we adopt the $8 \times 8$ UPA with the auxiliary beams and assume that both algorithms share the same AoD/AoA and estimated distance. Since POLAR only supports object sensing in 2D space, we further assume that it knows the elevation of targets in advance, which is used to transform the sensing result into 3D space. The results are shown in \figurename~\ref{fig:compare}, from which one could find that mm3NLoS performs better than POLAR regardless of whether the LoS path exists. The reason is that POLAR depends severely on the LoS path, and the sensing errors will be enlarged where only the NLoS path exists. Besides, POLAR is restricted to sense on the azimuth plane. When the azimuth mistakes of targets are relatively large, a non-negligible deviation will be caused. However, mm3NLoS can mitigate it by projecting the vectors to the other planes.

\section{Conclusion}
We proposed an mmWave enhanced sensing method named mm3NLoS, which could be used to address the performance degradation caused by the lack of LoS path. This method adopted the information of multiple NLoS paths (i.e., AoD, AoA, and ToF) to solve the problem from a geometric viewpoint. The simulation results demonstrated that the mm3NLoS algorithm could perform well under different scenarios and was not restricted by LoS conditions. Besides, it was worth mentioning that mm3NLoS also could be extended to multi-target sensing as long as the corresponding reflection paths of each target were estimated. To this end, hybrid antenna arrays could be adopted to sense diverse targets simultaneously. However, it was beyond the scope of this paper and would be investigated in future work.

\bibliographystyle{IEEEtran}
\bibliography{reference}

\end{document}